\documentclass[a4paper,11pt]{article}
\usepackage{aaskaiid}
\usepackage{orcidlink}
\usepackage{comment}

\title{Developments of the EVN to optimize SKA-VLBI and ngVLA synergies}

\author[1]{M. Kadler \orcidlink{0000-0001-5606-6154}}
\ShortName{Kadler et al.} 
\author[2]{S. Garrington}
\author[1,3,4]{F.~Eppel\orcidlink{0000-0001-7112-9942}}
\author[5]{F. Bertoldi}
\author[2]{R. Beswick}
\author[1,6,3]{A.-K. Baczko}
\author[7]{M. Brüggen}
\author[3]{A.~Brunthaler}
\author[8,9]{R. Deane}
\author[10]{D. Elsässer\orcidlink{0000-0001-6796-3205}}
\author[1]{C. M. Fromm\orcidlink{0000-0002-1827-1656}}
\author[3]{Y.~Y.~Kovalev\orcidlink{0000-0001-9303-3263}}
\author[3]{A. P. Lobanov}
\author[11,12,8]{J. Mc Kean}
\author[1]{K. Mannheim}
\author[13]{E.~J.~Murphy\orcidlink{0000-0001-7089-7325}}
\author[4]{Z.~Paragi\orcidlink{0000-0002-5195-335X}}
\author[2]{J.~F.~Radcliffe\orcidlink{0000-0002-0813-0497}}
\author[14]{D.~A.\ Riechers\orcidlink{0000-0001-9585-1462}}
\author[3]{E. Ros}
\author[15]{V. Rosero}
\author[13]{F. Schinzel}
\author[4,16]{A.~S\l{}owikowska
\orcidlink{0000-0003-4525-3178}}
\author[17,18]{L.~V.~Tóth\orcidlink{orcid:0000-0002-5310-4212}}
\author[19]{S. Vegetti}
\author[20]{P. de Vicente}
\author[21]{F. Walter}
\author[2]{N. Wrigley}
\author[22]{S. Wolf}
\author[1,3]{J. Wongphechauxsorn
\orcidlink{0000-0002-7730-4956}}
\author[3]{J. A. Zensus}
\affiliation[1]{Julius-Maximilians-Universit{\"a}t W{\"u}rzburg, Fakult{\"a}t für Physik und Astronomie, Institut für Theoretische Physik und Astrophysik, Lehrstuhl für Astronomie, Emil-Fischer-Str. 31, D-97074 W{\"u}rzburg, Germany}
\affiliation[2]{Jodrell Bank Centre for Astrophysics, Department of Physics and Astronomy, The University of Manchester, Manchester M13 9PL, UK}
\affiliation[3]{Max-Planck-Institut für Radioastronomie, Auf dem Hügel 69, 53121 Bonn, Germany}
\affiliation[4]{Joint Institute for VLBI ERIC, Oude Hoogeveensedijk~4, 7991 PD Dwingeloo, The Netherlands}
\affiliation[5]{Argelander Institut für Astronomie, Universität Bonn, Auf dem Hügel 71, D-53121 Bonn, Germany}
\affiliation[6]{Department of Physics and Astronomy, Chalmers University of Technology, SE-43992 Onsala, SE-41296 Gothenburg, Sweden}
\affiliation[7]{Hamburger Sternwarte, Universität Hamburg, Gojenbergsweg 112, D-21029, Hamburg, Germany}
\affiliation[8]{Department of Physics, University of Pretoria, Lynnwood Rd, Hatfield, Pretoria 002, South Africa}
\affiliation[9]{School of Physics, University of the Witwatersrand, 1 Jan Smuts Ave, Braamfontein, Johannesburg 2000, South Africa}
\affiliation[10]{TU Dortmund University, Department of Physics, Otto-Hahn-Str. 4, D-44227 Dortmund, Germany}
\affiliation[11]{Kapteyn Astronomical Institute, University of Groningen, Postbus 800, NL-9700 AV Groningen, the Netherlands}
\affiliation[12]{South African Radio Astronomy Observatory (SARAO), P.O. Box 443, Krugersdorp 1740, South Africa}
\affiliation[13]{National Radio Astronomy Observatory, 520 Edgemont Road, Charlottesville, VA 22903, USA}
\affiliation[14]{Institut f\"ur Astrophysik, Universit\"at zu K\"oln, Z\"ulpicher Straße 77, D-50937 K\"oln, Germany}
\affiliation[15]{Division of Physics, Mathematics, and Astronomy, California Institute of Technology, Pasadena, CA, 91125, USA}
\affiliation[16]{Institute of Astronomy, Faculty of Physics, Astronomy and Informatics, Nicolaus Copernicus University in Toru\'n, Grudzi\k{a}dzka~5, 87-100 Toru\'n, Poland}
\affiliation[17]{Institute of Physics and Astronomy, ELTE Eötvös Loránd University, H-1117 Budapest, Pázmány P. s. 1/A, Hungary}
\affiliation[18]{Institute of Physics, University of Debrecen, Hungary}
\affiliation[19]{Max Planck Institute for Astrophysics, Karl-Schwarzschild-Straße 1, D-85748 Garching bei München, Germany}
\affiliation[20]{Observatorio de Yebes, Instituto Geográfico Nacional, Cerro de la Palera, Yebes, Spain}
\affiliation[21]{Max Planck Institute for Astronomy, Königstuhl 17, D-69117 Heidelberg, Germany}
\affiliation[22]{Institut für Theoretische Physik und Astrophysik, Christian-Albrechts-Universität zu Kiel, Leibnizstraße 15, 24118, Kiel, Germany}
\abstract{
Operating in VLBI mode, the phased-up SKA-Mid will lend superior sensitivity
to global VLBI arrays, which opens the possibility to study the
milliarcsecond-scale physics of the faintest radio sources in the sky
with sensitivities significantly better than what can be achieved with
current arrays. At the upper frequency bands of SKA-Mid, between 1.2GHz
and 15GHz, strong synergies can be achieved with the upcoming ngVLA in
north America, if both new large facilities are connected through
compatible VLBI antennas in Europe.
To exploit the full science potential of these opportunities, we should
consider potential enhancements to the EVN 
by adding new antennas to improve spatial frequency
coverage. In addition, developments are needed toward increased
bandwidth, wider fields {of view}, new data-processing modes and correlation
hardware. The flexibility of EVN observations should also be improved
in  order to respond to evolving scientific demands, especially in the
context of transient and multimessenger astronomy. Here we discuss
possible expansions of the EVN through the addition of smaller
antennas/clusters including the 
LEVERAGE and RASOR concepts.
}

\begin{document}
\begingroup
\let\newpage\relax
\let\clearpage\relax
\maketitle
\endgroup
\section{Introduction}
The next decades of radio-astronomical research will be dominated by two large next-generation facilities: the Square Kilometre Array \citep[SKA; e.g.,][]{Braun2015} in Australia (SKA-Low) and South Africa (SKA-Mid) and
the next generation Very Large Array \citep[ngVLA;][]{Murphy2018} in North America, which will complement each other in their scientific capabilities.
In terms of frequency coverage, the SKA-Low/Mid will cover the range of (50--350)\,MHz and (0.35--15.3)\,GHz, respectively, while the ngVLA will operate from 1.2\,GHz to 116\,GHz. The baseline lengths between SKA-low and LOFAR (10-240\,MHz) or the Long Wavelength Array (LWA; 5-88\,MHz) are likely too large to be considered for many scientific applications at this frequency range, if combined through {Very Long Baseline Interferometry (VLBI)}. On the other hand, strong synergies can be developed in the range of mutual frequency coverage between SKA-Mid and the ngVLA from 1.2\,GHz to 15.3\,GHz. Being located on different continents, the SKA and ngVLA  will only have partial mutual sky visibility at any given moment in time but they can be combined through the technique of {VLBI} as long as an array of additional VLBI antennas can bridge the gap. 

Several science cases for both the SKA and the ngVLA  utilize  VLBI capabilities  \citep[][]{Paragi2015,Murphy2018,Kadler2023}. 
The superior sensitivity of the new infrastructures offers possibilities for groundbreaking new research in coordination with existing VLBI arrays. The European VLBI {Network} (EVN) is the preferred VLBI array for this purpose, given its geographic location, its large and sensitive single-dish antennas and the advanced infrastructure.
At the same time, it is also broadly understood that new developments and upgrades are needed in order to make full use of these new possibilities \citep[see, e.g.,][for the case of the  future of VLBI in Europe]{Venturi2020} in particular considering the development time line of the planned African VLBI network \citep{Gaylard2011,Atemkeng2020} and the heterogeneous design and stages of readiness of international VLBI arrays to co-observe with SKA phase-1 \citep{Rioja2023}.

\section{Current Status of SKA-VLBI/ngVLA-Long and Anticipated Capabilities}\label{sect:skavlbi_now}
In the first phase of the SKA-Mid project (often referred to as SKA\,1), the core of the SKA array will be phased up to operate as a highly sensitive element in global VLBI networks, a mode of operation referred to as {(SKA-VLBI)}. This will greatly enhance the overall sensitivity of those VLBI arrays but not strongly affect their angular resolution. In a sense, the SKA core can take on a role similar to the 26\,m antenna in Hartebeesthoek, which has been operated for many years successfully in observations with the {EVN} and the Australian Long Baseline Array (LBA). The largest improvement will be that the phased SKA core will strongly boost the baseline sensitivity on the longest baselines of both arrays and the image sensitivity. The biggest limitation to image fidelity in both cases will remain the wide $(u,v)$ gap resulting from the lack of VLBI telescopes in central Africa. This situation is discussed in more detail by \cite{Atemkeng2020,Gurvits2021} but not further addressed here. For the purpose of our study, we will assume that in addition to the SKA core, one remote antenna in South Africa (e.g., at the site of the Hartebeesthoek Observatory) and two additional remote antennas in neighboring countries will soon be available for VLBI in the same frequency bands, i.e., up to 15\,GHz, as offered by SKA1. For demonstration purposes we assume one antenna in Namibia and one in Botswana so that the four African VLBI sites would provide baseline spacings on the scale of a few hundred to a thousand kilometers. 
The LBA is currently the only VLBI network that is able to target the whole Southern sky. The main LBA array is located in Australia and New Zealand and the most frequently used transoceanic baselines have been provided in the past by inclusion of Hartebeesthoek. However, the LBA array has successfully been expanded with other transoceanic baselines to South America and Antarctica \citep[e.g.,][]{Ojha2010,Mueller2018} and to Asia. The EVN provides long North to South baselines between Europe and South Africa of up to about 9000\,km and North-East to South-West baselines of similar size into Asia.

Both the EVN and the LBA are heterogeneous arrays with only limited flexibility and agility in terms of transient responsiveness. For the SKA, on the other hand, transients will be an important key science topic. A flexible and sensitive subarray of antennas that can be triggered at short notice to provide long baselines to the North  would therefore be of great use. In recent years, the EVN has started to offer a subset of its telescopes for observations outside the regular EVN sessions (EVN lite), which is already finding increased use in transient follow-up and localization experiments \citep[e.g.][]{Hewitt2024}. 

SKA-VLBI is planned to enable the use of multiple simultaneous tied-array beams, which puts further constraints on the technology with which  future AVN antennas need to be outfitted and also on technology upgrades on the existing stations of the EVN and LBA \citep{Rioja2023}. Specifically, multi-pixel technology or MultiView techniques need to be enabled, which is already built into the design of the ngVLA Long multi-antenna concept. 
In the currently planned design, the ngVLA Long will contain a subarray of thirty eighteen-meter antennas (ten stations of 2--4 antennas, each) distributed across North America with a maximum baseline length of about 9000\,km. It will be possible to combine this \textsl{ngVLA Long} subarray with antennas from the \textsl{ngVLA Mid} array to improve the $(u,v)$ coverage at short baselines and to enhance sensitivity. Details of the ngVLA Long design are discussed in the ngVLA memo series\footnote{See \href{https://ngvla.nrao.edu/page/memos}{ngVLA Memos} \#84, \#105.}. 

\section{Leverage from Europe for next-generation VLBI science}
Antennas in Europe are in a geographically favorable location to co-observe both with the ngVLA and the SKA in VLBI mode to enhance the synergy between these {two next generation VLBI (ngVLBI) facilities}. Here, we describe two concepts under consideration. Both are based on the idea of advancing the capabilities of the EVN through (clusters of) modern, fast and flexible radio telescopes of moderate diameter building on existing infrastructure and technology design.

\subsection{The LEVERAGE concept}
LEVERAGE (Long-baseline Extension in next-generation VLBI Experiments and Rapid-response Array Germany) is a proposed network of mid-to-high-frequency antennas in Germany, with the potential for expansion across Europe and implementation into the EVN \citep{Kadler2024}.
The array is composed of several clusters of antennas, separated by distances of several hundred kilometers.
LEVERAGE serves a dual purpose: it complements the ngVLA and SKA-VLBI networks while also functioning as a stand-alone facility for transient studies and other scientific applications, thereby strengthening the EVN's capabilities in multimessenger astronomy.
LEVERAGE is designed to  enhance the $(u,v)$ coverage beyond 5000\,km in the ngVLA Long subarray and to connect  to African telescopes (Bempong-Manful et al, this volume), improving image fidelity in SKA-VLBI observations for the upper SKA-Mid frequency bands.

A larger number of antennas per station offers the possibility of enhancing the sensitivity, which is of particular importance on the longest baselines. For single antenna stations, the technique of source switching can be used for phase referencing (which reduces the effective on-source time), while multiple antennas in one station open a range of paired antenna-calibration schemes.  
At this point, the exact number of \textsl{LEVERAGE} stations, as well as the number of telescopes per station and the layout of the array are still subject to further planning, tests and simulations.
For the purpose of this work, we consider four stations in Germany: one near Effelsberg, one in Saxony Anhalt, one in Saxony, and one in Bavaria (Alps), each equipped with four telescopes per station.

The  SKA-Mid and ngVLA synergy frequency band extends from 1.2\,GHz up to 15.3\,GHz, which thus will be a priority regime for \textsl{LEVERAGE} receivers. In general, it would be desirable that a majority of EVN-station VLBI backends  support the SKA-Mid  bands 1, 2 and 5a/b (phase-1), 3, 4, and 6 (phase-2), as well as the ngVLA bands 1--3.

\subsubsection{LEVERAGE for snapshot VLBI observations with the ngVLA and SKA-VLBI}\label{sect:transients}
In many types of transient radio sources, the associated radio emission is  close to the sensitivity limits of the current generation of radio telescopes. In spite of this, impressive results could already be derived \citep[e.g.,][]{Corsi2018a,Bright2023}, raising even greater hopes for the hypersensitive next-generation facilities. The need for radio follow-up observations of faint sources in the mJy or $\mu$Jy regime is rapidly growing in the next decade with the increasing importance of large time-domain surveys such as the Rubin Observatory Legacy Survey of Space and Time \citep[LSST;][]{Li2022} and later because of the great promise of multimessenger astronomy in coordination with  gravitational-wave astrophysics when the Laser Interferometer Space Antenna \citep[LISA; ][]{Amaro-Seoane2017} is in operation.

Already with current instruments, VLBI  is an underused technique  in the context of radio-transient research \citep[e.g.,][]{Sokolovsky2018}. The high angular resolution offered by  VLBI can allow us to measure source positions, sizes and proper motions  with sub-milliarcsecond precision, detect polarized emission from spatially resolved regions that are hidden by in-beam depolarization for lower-resolution instruments and detect faint compact transient sources against brighter diffuse emission regions thanks to the spatial-filter properties of VLBI.  
In explosions and other transient events, high frequencies 
can generally probe ultra-compact emission regions which are optically thick at longer wavelengths due to self absorption. 

Specific examples of important science cases that require VLBI snapshot capability are listed, e.g.,
in \cite{Aharonian2013} and \cite{Kadler2023} and include compact-object mergers and their associated electromagnetic emission \citep[e.g.,][]{Corsi2018b,Ravi2018}, fast radio bursts \citep[e.g.,][]{Marcote2017,Platts2019}, stellar transients \citep[e.g.][]{Marcaide2009,Mueller2023,Palliyaguru2021}, gamma-ray bursts \citep[GRBs; e.g.][]{Kumar2015,Salafia2022}, as well as high-energy gamma-ray and neutrino flares in AGN jets, supernovae or GRBs \citep[e.g.][]{Ros2020,Abbasi2023,Stein2021}.

 For such applications the snapshot observing mode of VLBI facilities is crucial.
Snapshots are short observations up to a few minutes in duration. The resulting data quality is limited by the sidelobe level, i.e., the density of coverage in the $(u,v)$ plane, which is considerably inferior for snapshots compared to longer tracks. Consequently, the snapshot mode is most attractive for studies of bright and compact sources, e.g., to characterize and monitor their brightness temperatures, for studies in the time domain and for astrometry. 
It has been shown that brightness temperatures can be effectively constrained, even in cases of sparse $(u,v)$ coverage or, in the extreme case, with single baselines \citep{Lobanov2015}. However, the effectiveness of this approach depends on the extent and uniformity of the array and thus changes substantially over one day due to Earth rotation.

The system requirements that have guided the ngVLA design included a snapshot-fidelity requirement as follows: \textsl{``The ngVLA snapshot performance should yield high fidelity imaging on angular scales $>100$\,mas at 20 GHz for strong sources''}. Consequently, the configuration of antenna locations in the ngVLA Main subarray has been optimized accordingly but the snapshot fidelity on more compact scales provided by the ngVLA Long has weaknesses, namely a much wider East-West extent of the Long array compared to its North-South dimensions and sparse coverage in the outer parts of the $(u,v)$ plane compared to the inner regime. Coverage at short $(u,v)$ spacings can be enhanced by additional antennas of the ngVLA Mid subarray coobserving with the Long array. 

\begin{figure*}[htb]
    \centering    
    \includegraphics[width=0.32\textwidth]{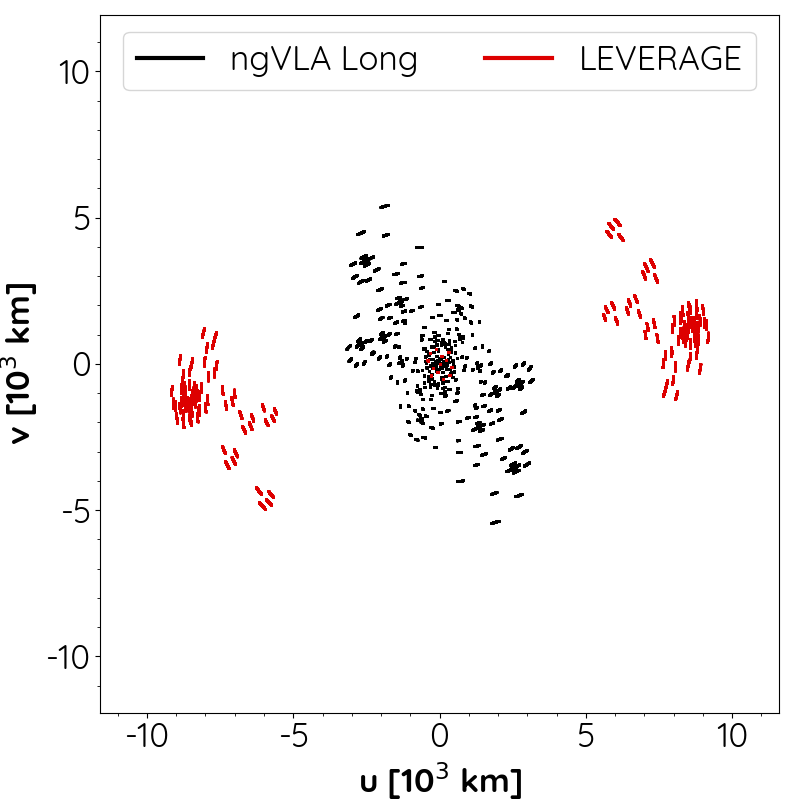}
    \includegraphics[width=0.32\textwidth]{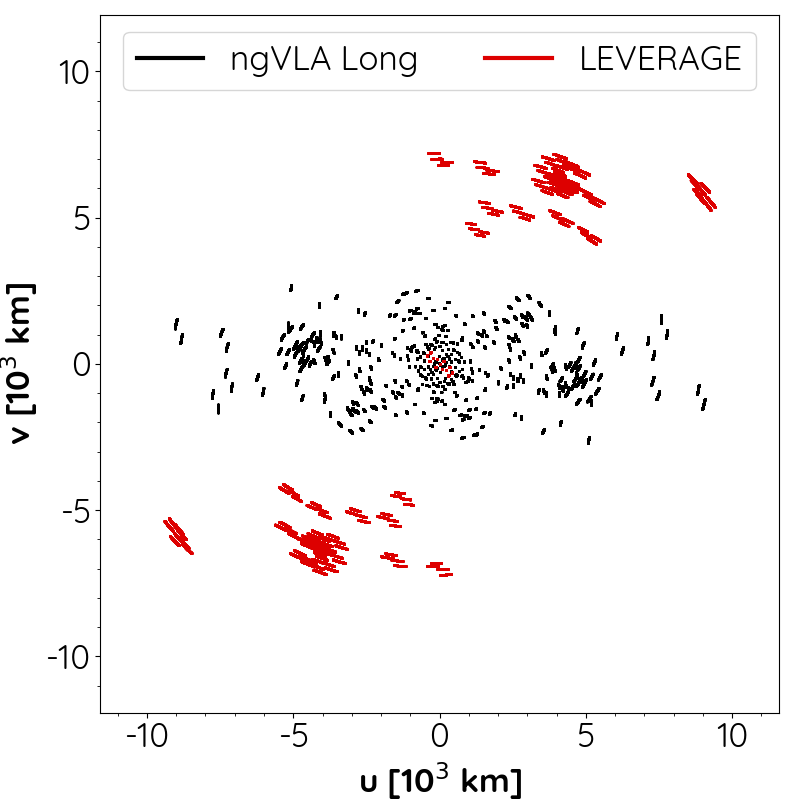}
    \includegraphics[width=0.32\textwidth]{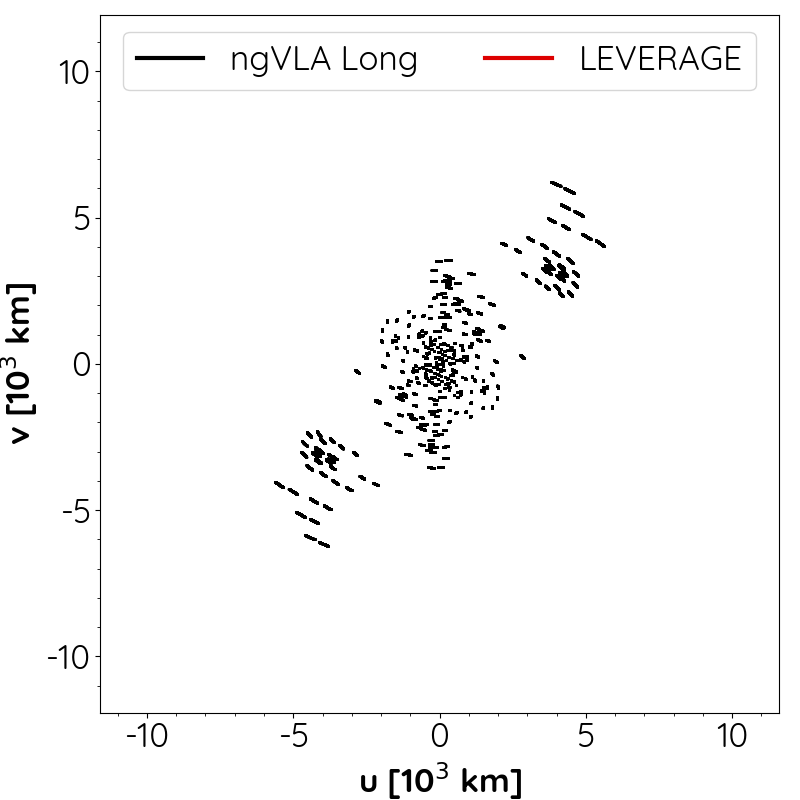}
    \includegraphics[width=0.32\textwidth]{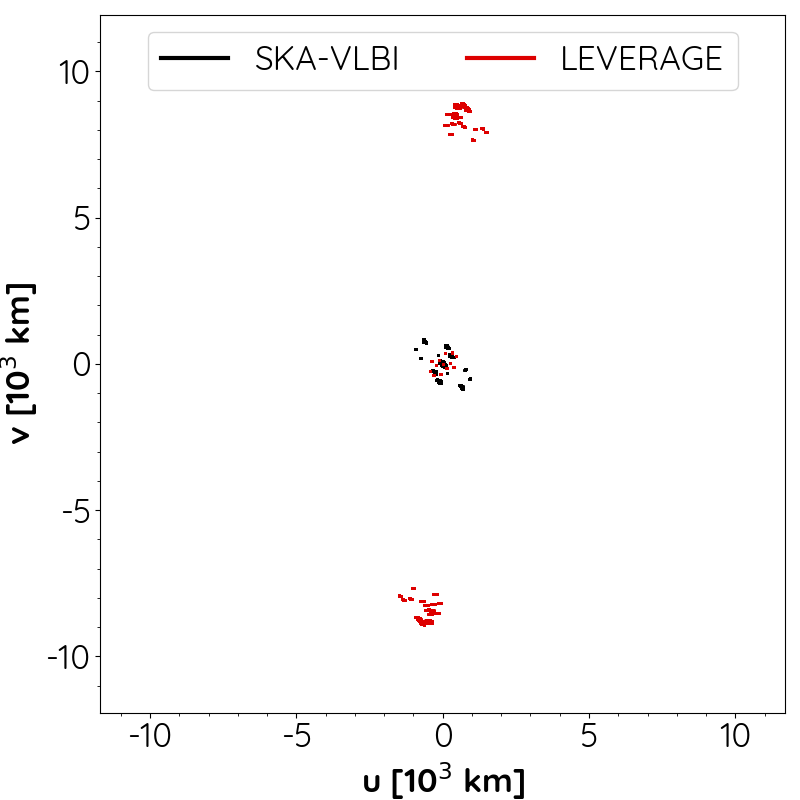}
    \includegraphics[width=0.32\textwidth]{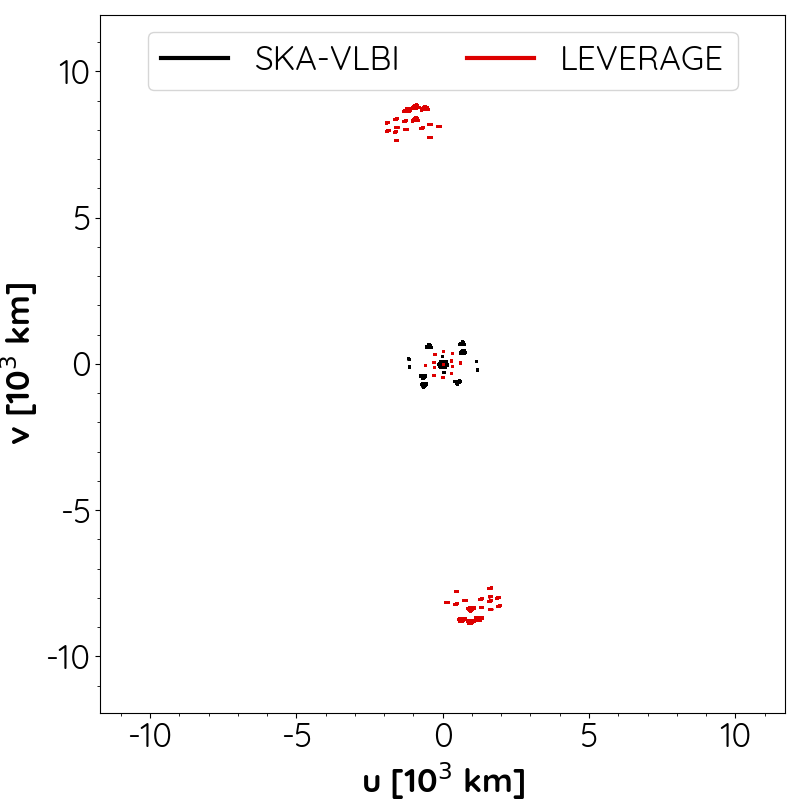}
    \includegraphics[width=0.32\textwidth]{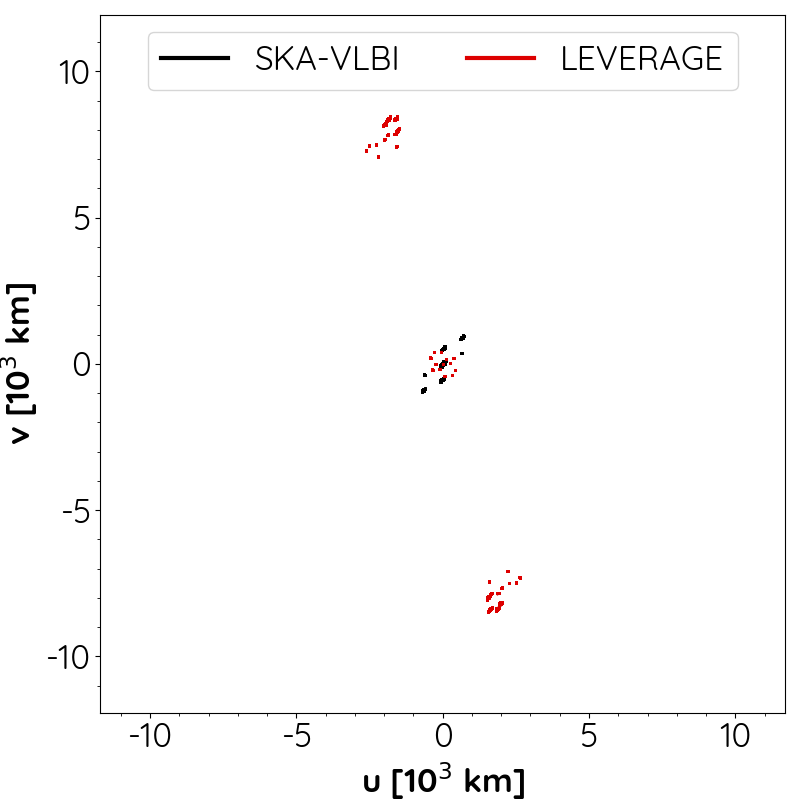}
    \caption{Top: Three example $(u,v)$ coverages for 12\,min snapshot observations with the ngVLA Long in its current design (black dots) for a source at declination 44$^\circ$ at the three different times marked with red vertical lines in the top panel of Fig.~\ref{fig:uptime44} to illustrate the changing snapshot response of the array due to Earth rotation. The additional coverage provided by four LEVERAGE stations in Europe (Germany) is shown as red dots. Bottom: Three example $(u,v)$ coverages for 12\,min snapshot observations with the four selected African SKA-VLBI stations (black dots) for a source at declination 18$^\circ$ at the three different times marked with red vertical lines in the bottom panel of Fig.~\ref{fig:uptime44}  to illustrate the changing snapshot response of the array due to Earth rotation. The additional coverage provided by four lever stations in Europe (Germany) is shown as red dots.}
    \label{fig:44deg_snapshots}
\end{figure*}

\begin{figure*}[htb]
    \centering
        \includegraphics[width=0.49\columnwidth]{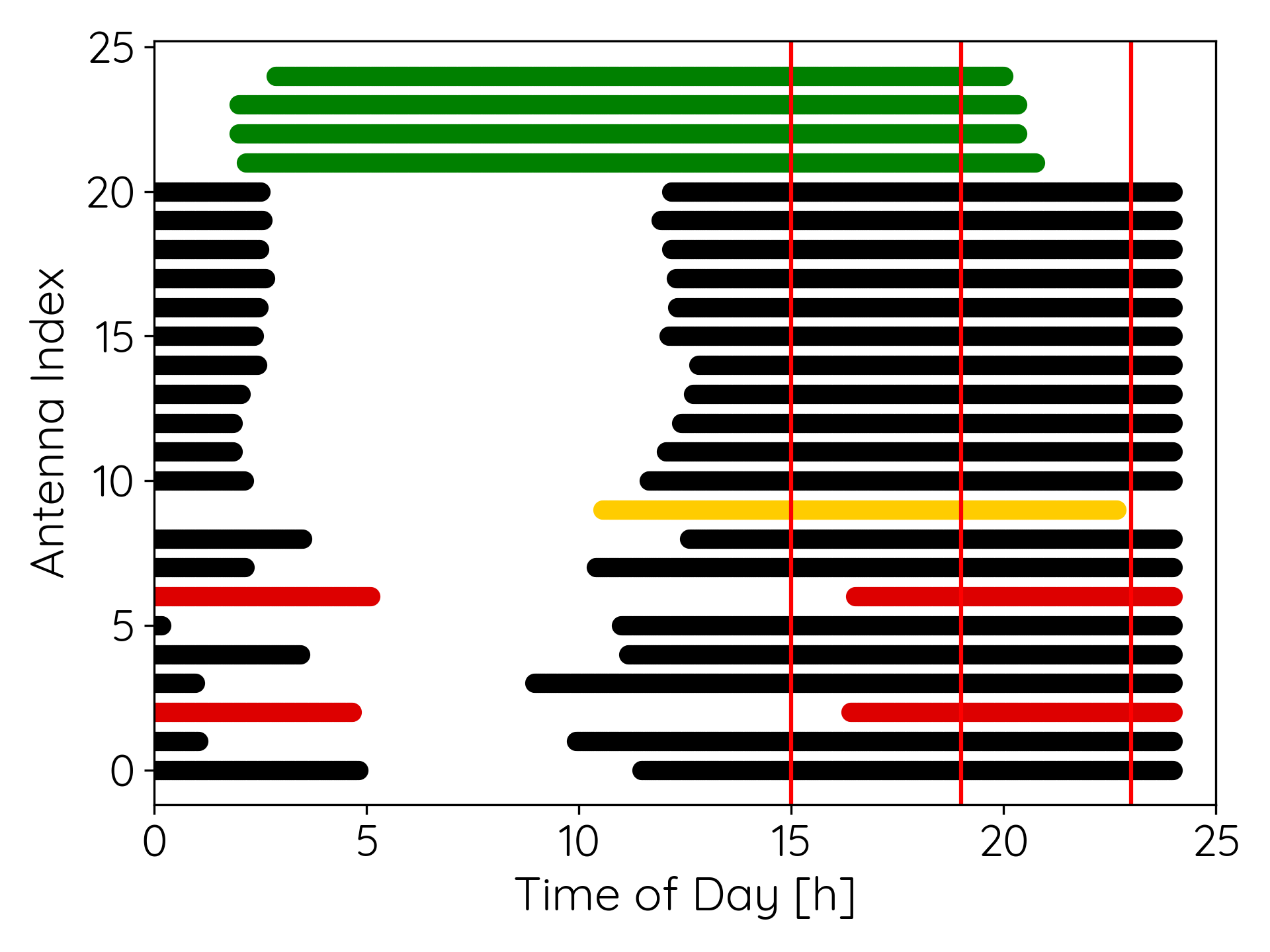}
        \includegraphics[width=0.49\columnwidth]{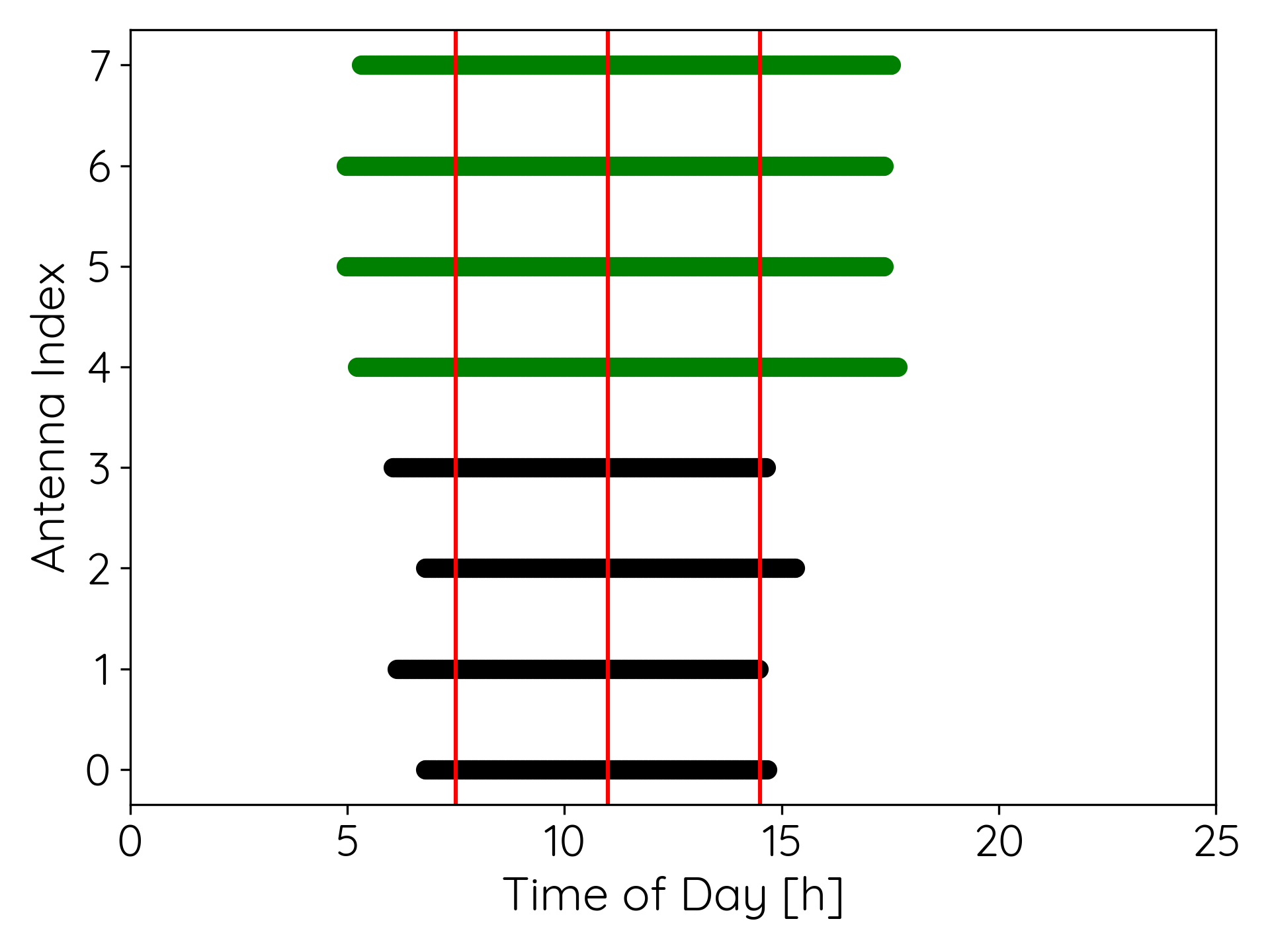}
   \caption{\textsl{Left:} Uptime plot showing the visibility of a source at 44$^\circ$ declination for the ngVLA Long/Mid continental antennas (black), two antennas on Hawaiian islands (red), a Puerto Rican antenna (yellow) and four antennas in Germany (green). Red vertical lines mark three different times for which the top panels of Fig.~\ref{fig:44deg_snapshots} shows the corresponding snapshot $(u,v)$ coverages. \textsl{Right:} Uptime plot showing the visibility of a source at 18$^\circ$ declination for african VLBI stations (black) and four antennas in Germany (green). Red vertical lines mark three different times for which the bottom panels of Fig.~\ref{fig:44deg_snapshots} shows the corresponding snapshot $(u,v)$ coverages.}
    \label{fig:uptime44}
\end{figure*}

In Fig~\ref{fig:44deg_snapshots}, we show three snapshot $(u,v)$ coverages of the ngVLA (black dots; 10 ngVLA Long stations plus eight ngVLA antennas from the central array at the tips of the ngVLA Mid spiral  and the old VLBA sites; see \href{https://ngvla.nrao.edu/page/memos}{ngVLA Memo \#105}) for three different times and assuming an observation at  declination 44$^\circ$. The three snapshot times correspond to times when the source is \textsc{i)} not yet visible for the ngVLA stations on Hawaii, \textsc{ii)} observable for the full ngVLA Long array, and \textsc{iii)} the source has already set for the Puerto Rico station (see Fig~\ref{fig:uptime44}). Evidently, the $(u,v)$ coverage is best in case \textsc{ii)} and the resulting beam shapes of the array vary substantially. 
We then add additional stations in central Europe to investigate how this can improve the coverage on the longest baselines. 
The red dots in Fig.~\ref{fig:44deg_snapshots} show the strong influence that these  additional European stations  provide in cases \textsc{i)} and \textsc{ii)}, while they could not add to the coverage in case \textsc{iii)}. Inspection of  Fig.~\ref{fig:uptime44} shows that at 44$^\circ$ declination, the period of nonvisibility to the European stations is short ($\sim 5$\,hours). This gap gets longer for declinations further South and smaller for more Northern targets.

Rapid ($<1$\,day) follow-up radio observations are key to a proper modeling and understanding of radio transients 
{\cite[see, e.g.,][for the cases of short and long gamma-ray bursts, respectively]{Lloyd-Ronning2018,Bright2023}}.
For any given rapidly evolving radio transient in the northern sky, the chances that the ngVLA  can immediately observe it are only about 50\,\% ($\sim$ 8--14 hours source visibility depending on declination)  given the elevation limits at the various sites. Even during the visibility window, the properties of the array will change considerably because of Earth-rotation-induced baseline projection effects. The most drastic change happens when the target source rises above the horizon for the Hawaii antennas or drops below the horizon at Puerto Rico, which immediately introduces a jump in maximum baseline and thus in resolving power by up to $\sim 50$\,\%. 
This limits the time per day that any given source can be monitored with a uniform brightness-temperature threshold. For any science case that needs this capability, such as the monitoring of rapidly evolving radio-transient sources, a more uniformly evolving snapshot reception pattern would be desirable.

\begin{figure*}
    \centering
    \includegraphics[width=0.85\linewidth]{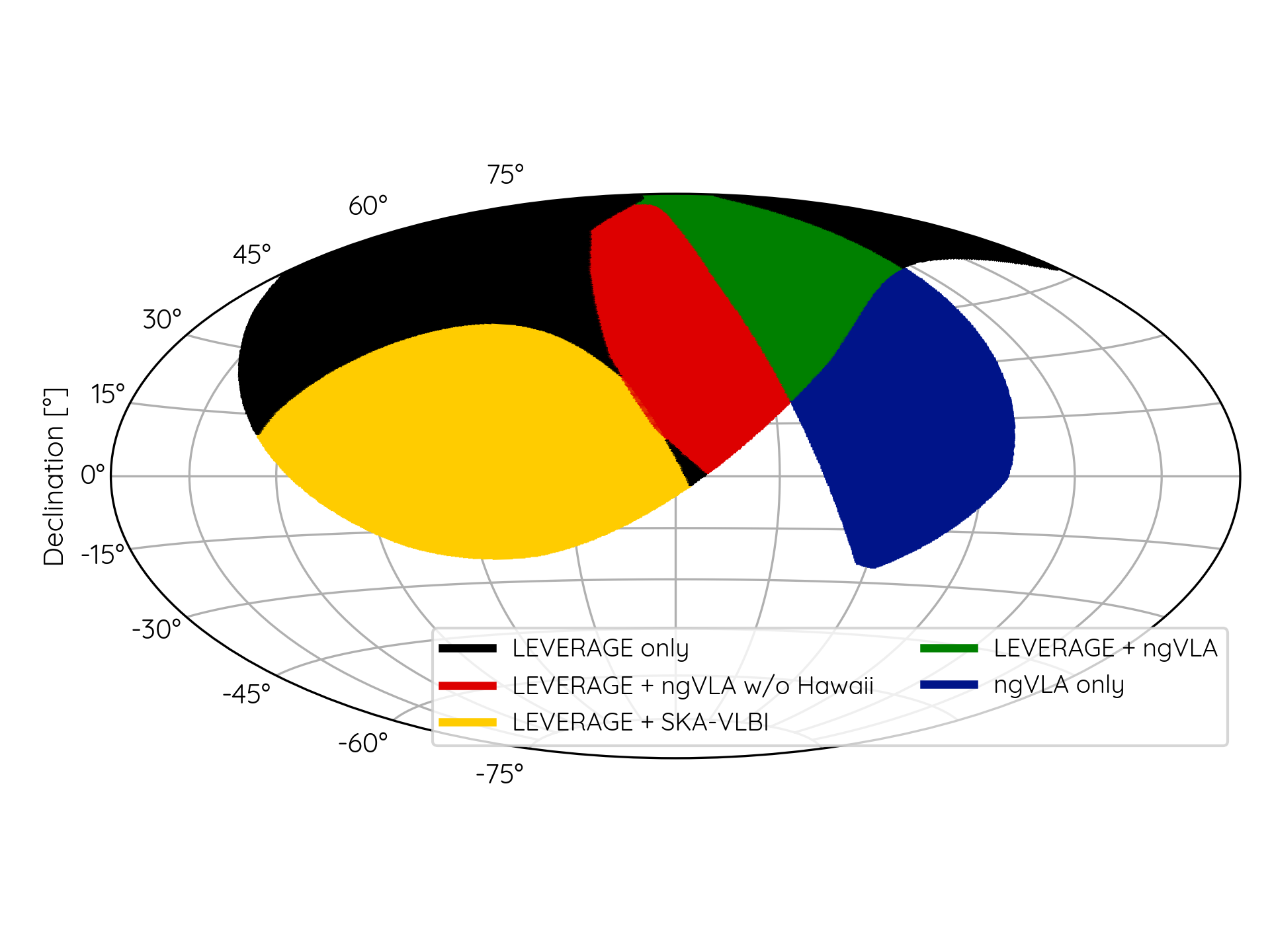}\hfill
    \caption{Fraction of the full sky (at an arbitrary time) in which  transient events can be followed up immediately with $>5000$\,km baselines with the ngVLA only, with the SKA only and with a \textsl{LEVERAGE} array from Europe. }
    \label{fig:snapshotfraction}
\end{figure*}

Figure~\ref{fig:snapshotfraction} shows the instantaneous responsiveness with a $>5000$\,km baseline snapshot observation to transient alerts at an arbitrary time for future VLBI arrays that include the ultrasensitive ngVLA and/or the SKA.
 At typical northern declinations, 20\,\% to 40\,\% of all events can be traced if the \textsl{LEVERAGE} of the proposed additional European antennas is included, roughly doubling the number of traceable events for the ngVLA Long, alone. Near the equator and at moderate negative declinations, the coverage including the SKA is nearly fully complementary to the ngVLA.

\subsubsection{LEVERAGE for full-track ngVLA-Long and SKA-VLBI observations}
\begin{figure}
    \centering
    \hspace*{0.1cm}\includegraphics[width=\linewidth]{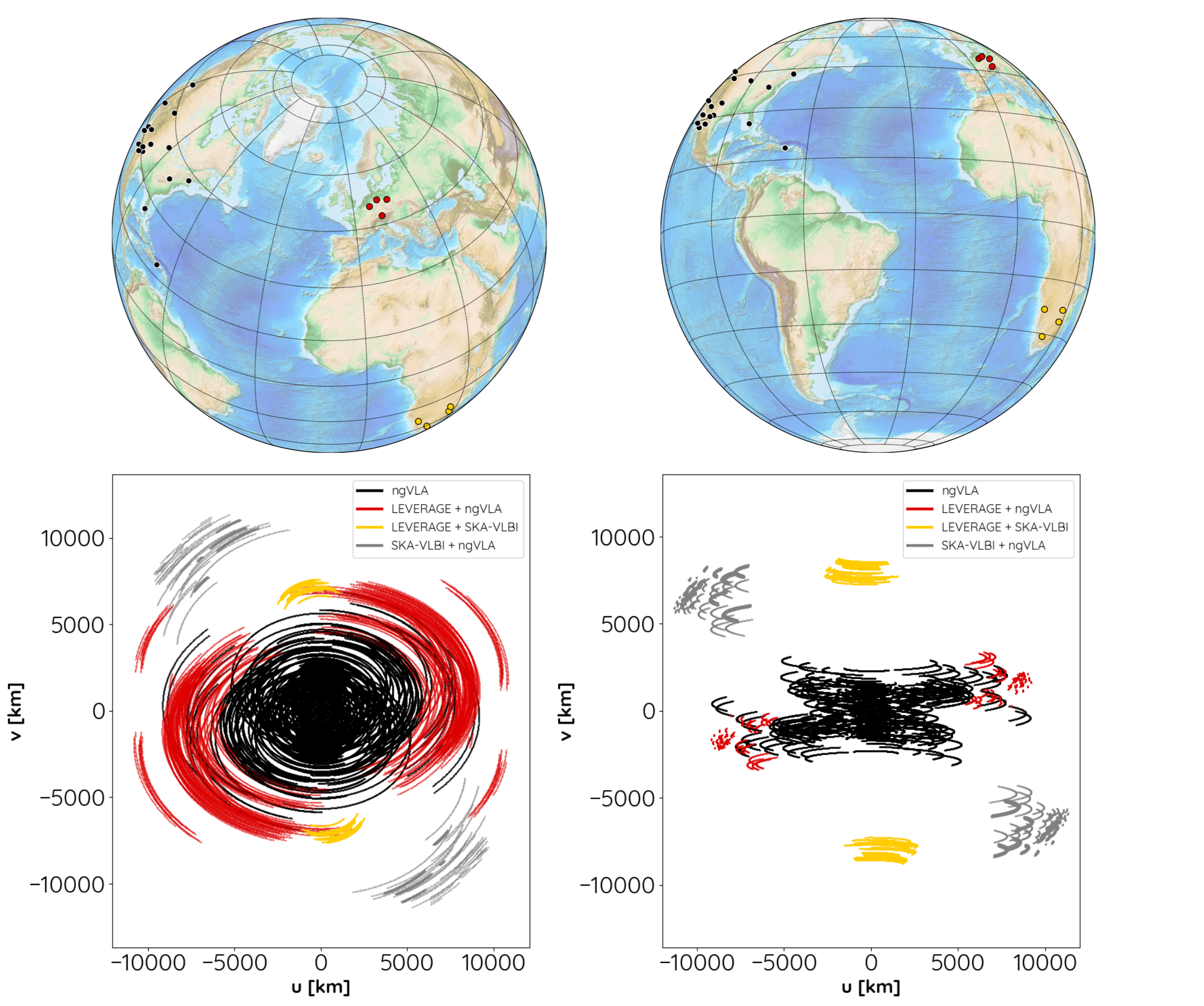}
    \caption{Top: Distribution of ngVLA Long and \textsl{LEVERAGE} antennas on the surface of the Earth as seen from a geostationary orbit at $+44^\circ$ (left) and at $-6^\circ$ (right) declinations. Bottom: $(u,v)$ coverages plot for full-track observations with the ngVLA Long (black dots) and the additional coverage provided by \textsl{LEVERAGE} in combination with the ngVLA (red dots) and with African SKA-VLBI stations (yellow and gray dots) for sources at the two declinations.}
    \label{fig:uv_fulltracks}
\end{figure}

Figure~\ref{fig:44deg_snapshots} and Fig.~\ref{fig:uptime44} already demonstrate that complementary antennas in Europe improve the $(u,v)$ coverage of the ngVLA Long significantly on the longest spacings and at high declinations.  Far further south, source mutual visibility from mid Europe rapidly decreases because of the high latitude and large longitude difference. Furthermore, as discussed in \href{https://ngvla.nrao.edu/page/memos}{ngVLA Memo \#84}, it is beneficial if Long stations are not left free to float in the calibration process. In this respect, it helps a lot if each station has a partner station at distances of a few hundred kilometers.
The actual distribution of these additional stations does not have a large impact on the overall $(u,v)$ coverage.  We choose again the four stations in Germany to demonstrate the impact that such a \textsl{LEVERAGE} array can have.
Germany is about seven to fifteen degrees west of the SKA core site, which enables interesting options for synergies between the ngVLA Long and SKA-VLBI. In Fig.~\ref{fig:uv_fulltracks}, we demonstrate the improved $(u,v)$ coverage that can be achieved across a broad range of declinations between the ngVLA Long and the four assumed SKA-VLBI sites (see Sect.~\ref{sect:skavlbi_now}). The large gap between the USA and South Africa is bridged by the European stations. At Northern declinations, e.g. at $44^\circ$, the additional baselines to the SKA-VLBI stations provides extended $(u,v)$ coverage on very long baselines in the NW--SE direction. At lower declinations closer to the equator, which spans the largest fraction of sky area, the European stations provide North-South long-baseline coverage, extending the ngVLA Long to SKA-VLBI baselines in this part of the $(u,v)$ plane. Note that the remaining gap between 2500\,km and 7500\,km in the North-South direction can be closed by additional AVN antennas in central Africa \citep[see][and Bempong-Manful et al., this volume]{Atemkeng2020,Bempong-Manful01.2026.SKA}.

\subsubsection{The Wetterstein Millimeter Telescope (WMT)}
The Wetterstein Millimeter Telescope {\citep[WMT; ][]{Kadler2024,Neidhardt2026}} is a planned radio telescope associated with the  Environmental Research Station (UFS) Schneefernerhaus.
The WMT is envisioned as a state-of-the-art instrument, playing a critical role in Germany's contribution to the ngVLA and European VLBI efforts. As the pioneering element of the \textsl{LEVERAGE} concept, the WMT ultimately aims to enhance the geographic coverage in East-West baselines with the ngVLA, and in  North-South baselines with SKA-mid \citep{Kadler2024}. 

At 2650\,m, about 300\,m below the summit of the Zugspitze, which is the highest peak of the Wetterstein Mountains and the highest mountain in Germany, the UFS is Germany's highest-altitude research station.  
The UFS site provides excellent conditions for millimeter-wave observations as demonstrated by 
an extensive database on atmospheric measurements which have been compiled over decades. Moreover, the site provides  access to the UFS infrastructure for power, data, and maintenance and promises close links to be established with environment-related research fields. A comprehensive overview of the present scientific activities conducted at the Schneefernerhaus is presented by \cite{dlr189714}. 

The WMT has been designed to work on a wide range of astrophysical topics, many of them aligned with the goals of the \textsl{LEVERAGE} concept. These topics
include black holes and their relativistic jets, protoplanetary disks, dark matter and galaxy evolution.
In addition to astrophysics, the WMT science program extends to interdisciplinary research cases, such as the study of near-Earth asteroids and small bodies in the solar system using planetary radar technology, atmospheric physics studies and satellite communication.

\subsection{The RASOR concept}
The e-MERLIN array in the United Kingdom already provides baselines from 10 to 200\,km and operates in 3 bands (1.25-1.75, 4-8 and 18-25 GHz). It is frequently combined with the EVN to provide baselines from 10--10,000km. 

The Radio Astronomy and Space Observation Research facility (RASOR) has been developed by a team in the United Kingdom including collaborators from the Universities of Manchester, Birmingham, Bath and Oxford along with the STFC RAL Space Weather group. RASOR is designed to offer powerful capabilities for radio astronomy, space weather and space situational awareness. It would extend e-MERLIN with a number of SKA-mid (or similar) dishes and LOFAR HBA (or similar) stations spread across the whole UK, with up to 850\,km maximum baselines. The low frequency stations are primarily designed for space-weather observations, while the mid frequency dishes would be used for radio astronomy observations and as receivers for multi-static radar observations of satellites and debris in earth orbit. The current RASOR proposal includes approximately 10 new SKA dishes and 4 new LOFAR stations. The precise numbers would depend on funding availability: the RASOR proposal was very strongly supported by STFC in the UK, but is not yet funded. 

 The combination of existing EVN sites with new RASOR and potential other new cm-wavelength antennas such as LEVERAGE at existing LOFAR stations for example would transform the aperture coverage and imaging fidelity of the EVN. Crucially the availability of a significant number of dedicated VLBI dishes would also transform the capabilities of the EVN for rapid response science, enabling a subset of antennas to be available for observations at any time. 

This enhanced EVN would be a natural partner for SKA-VLBI observations, given the common longitude. Assuming an elevation limit of 15 degrees for both RASOR and SKA-mid dishes the accessible (equatorial) declination range for joint VLBI observations would be more than 60 degrees (-22,+45).

\section{Summary and Conclusions}
An important milestone towards SKA-VLBI was achieved in the first half of 2025 with the successful execution of the first joint scientific VLBI observations using the EVN and MeerKAT, with the data correlated at the Joint Institute for VLBI ERIC \citep{EVN-MeerKAT-2025}\footnote{Press release on behalf of SARAO and JIV-ERIC Team: \textit{An Earth-sized radio observatory just got better: South Africa's MeerKAT telescope joins forces with the European VLBI Network of telescopes}, \url{https://www.jive.eu/news/earth-sized-radio-observatory-just-got-better-south-africas-meerkat-telescope-joins-forces}}. These observations demonstrated the substantial gains in sensitivity and imaging capability enabled by integrating MeerKAT into the EVN, thereby serving as a pathfinder for future SKA-VLBI operations. Building on this success, from 1st of October 2025, MeerKAT has been offered as part of the regular EVN Call for Proposals in L band for approximately 100 observing hours per year. The community has recognised this new capability, and the EVN Programme Committee has received multiple proposals requesting MeerKAT participation, highlighting the scientific demand for this enhanced capability.

In this chapter, we have discussed
the current design and science capabilities of the ngVLA Long and the current status of SKA-VLBI plans, as well as the limitations of both facilities in terms of high-resolution image fidelity and responsiveness to transient events.  We introduced two concepts as proposed future expansions of the EVN to optimize synergies with these next-generation arrays. 

Both the LEVERAGE and RASOR concepts provide independent networks of modern radio telescopes.
While  high-impact science is to be expected from both facilities also in stand-alone observing mode, we have focused here on the  improvements these facilities can provide to 
next-generation VLBI observations for sub-milliarcsecond-scale high-fidelity imaging.
The science cases to be addressed by both concepts and resulting design considerations do align very well with the EVN Science Vision document \textsl{VLBI20-30: a scientific roadmap for the next decade} \citep{Venturi2020}. Specifically, the development of SKA-VLBI and  high-frequency capabilities at increased bandwidths are important goals of European radio astronomy.

{\bf Acknowledgement}:
 We acknowledge support by the German Federal Ministry of Education and Research (BMBF) under ErUM-Pro grant 05A23WW3 (Verbundprojekt D-MeerKAT III).

\bibliographystyle{abbrvnat-maxbibnames4}
\setlength{\bibsep}{0.0pt} 
\newcommand{\actaa}{Acta Astron.} 
\newcommand{\araa}{ARA\&A} 
\newcommand{\aar}{A\&ARv} 
\newcommand{\aapr}{A\&ARv} 
\newcommand{\ab}{Astrobiol.} 
\newcommand{\aj}{AJ} 
\newcommand{\apj}{ApJ} 
\newcommand{\apjl}{ApJL} 
\newcommand{\apjs}{ApJSS} 
\newcommand{\ao}{Appl. Opt.} 
\newcommand{\apss}{Astro. \& Space Sci.} 
\newcommand{\aap}{A\&A} 
\newcommand{\aaps}{A\&AS.} 
\newcommand{\baas}{Bull. Am. Astron. Soc.} 
\newcommand{\caa}{Chinese A\&A} 
\newcommand{\cjaa}{Chinese J. A\&A} 
\newcommand{\cqg}{Class. Quantum Gravity} 
\newcommand{\gal}{Galaxies} 
\newcommand{\gca}{Geo. Cosmo. Acta} 
\newcommand{\icarus}{Icarus} 
\newcommand{\jcap}{JCAP} 
\newcommand{\jgr}{J. Geophys. Res.} 
\newcommand{\jgrp}{J. Geophys. Res. Planets} 
\newcommand{\jqsrt}{J. Quant. Spectrosc. Radiat. Transf.} 
\newcommand{\memsai}{Mem. SAIt} 
\newcommand{\mnras}{MNRAS} 
\newcommand{\nat}{Nature} 
\newcommand{\nastro}{Nat. Astron.} 
\newcommand{\ncomms}{Nat. Commun.} 
\newcommand{\nphys}{Nat. Phys.} 
\newcommand{\na}{New Astron.} 
\newcommand{\nar}{New Astron. Rev.} 
\newcommand{\physrep}{Phys. Rep.} 
\newcommand{\pra}{Phys. Rev. A} 
\newcommand{\prb}{Phys. Rev. B} 
\newcommand{\prc}{Phys. Rev. C} 
\newcommand{\prd}{Phys. Rev. D} 
\newcommand{\pre}{Phys. Rev. E} 
\newcommand{\prx}{Phys. Rev. X} 
\newcommand{\prl}{Phys. Rev. Let.} 
\newcommand{\psj}{Planet. Sci. J.} 
\newcommand{\planss}{Planet. Space Sci.} 
\newcommand{\pnas}{Proc. Natl Acad. Sci. USA} 
\newcommand{\procspie}{Proc. SPIE} 
\newcommand{\pasa}{PASA} 
\newcommand{\pasj}{PASJ} 
\newcommand{\pasp}{PASP} 
\newcommand{\rmxaa}{RMXAA} 
\newcommand{\sci}{Science} 
\newcommand{\sciadv}{Sci. Adv.} 
\newcommand{\solphys}{Sol. Phys.} 
\newcommand{\sovast}{Soviet Ast.} 
\newcommand{\ssr}{Space Sci. Rev.} 
\newcommand{\uni}{Universe} 

\bibliography{EVN-SKA-VLBI}

\end{document}